Translating photobiology to electrophysiology

A brief overview of several photobiological processes with accent on electrophysiology

Vadim Volkov

Faculty of Life Sciences and Computing, London Metropolitan University, 166-220 Holloway Road, London N7 8DB, UK

For correspondence: v.volkov@londonmet.ac.uk and vadim.s.volkov@gmail.com

The mini-review gives special attention to holistic approach and mechanisms of processes. The physical and chemical frames and background for visual perception and signalling are discussed. Perception of photons by retinal rod cells is described in more detail starting from photon absorption and culminating in ion currents. Dark noise and temperature-dependence of photoreceptor cells are analysed. Perception of polarised light, its effects and informational importance are discussed based on underlying mechanisms and specialised morphological structures of biological organisms. Role of statistics of photons in photoreception is questioned. The review also pinpoints new and developing directions and raises questions for the future research.

Photobiology – electrophysiology – polarised light – rod cells – dark noise – statistics of photons - optogenetics

Life has evolved, developed and is flourishing nowadays in the permanent fluctuating fluxes of electromagnetic radiation of different frequencies and amplitudes. The electromagnetic waves are bringing information about the world and provide energy for the existence of living organisms. From the wide range of electromagnetic waves life chose narrow band of 400-800 nm for vision, light perception and photosynthesis. The reasons are in energy of photons and nature of chemical bonds. Much higher energies are harmful and break molecules; much lower energies are not distinguished from the thermal noise (see examples below).

The short review is focused on physico-chemical mechanisms of photobiological processes, conversion of photons into ion currents for further processing and on more specific aspects including perception of polarised light and light with unusual statistics, their role for biological systems and implications for biological research. The general preface starts from storage of energy in photosynthesis moving to mechanisms of light perception and vision.

Quanta of light (photons) are caught by pigments (usually chlorophylls) in photosynthetic membranes of plants, algae and some bacteria; the sequence of fast picosecond photochemical reactions results in electric charge transfer between the sides of the membranes. The generated electric potential is used for downhill charge transfer; the energy is converted to energy of chemical bonds, mostly in ATP (adenosine triphosphate), so the energy can be stored for further biochemical reactions and membrane transport processes (Govindjee, 1982). The complex chain of reactions for ATP synthesis requires large protein macromolecular complexes. Usually protons are passing via subunits of ATP-synthase to change the conformation of protein complex; finally it is released in ATP synthesis from ADP and inorganic phosphate $P_i$ (Boyer, 1997; McCarty et al., 2000).

Energy E of photon (quantum of light) is determined by the equation:

$$E = h \nu = h c / \lambda ,$$

where $\nu$ is the light wave frequency, which is reverse proportional to wavelength $\lambda$ ($c = \lambda * \nu$), h is Planck's constant (about $6.6 * 10^{-34}$ J s) and c is light speed in vacuum (about $3*10^8$ m/s). The energies of single visible photons (400 – 700 nm, see below) will be then from 2,8 to 5 $10^{-19}$ J (reverse proportional to wavelength). The values correspond to 170-300 kJ per mole of photons. For comparison, the standard energy of ATP hydrolysis is -30 kJ/mole and up to around -60 kJ/mole depending on pH, ATP, ADP and ion concentrations (Alberty, 1969; Kammermeier et al., 1982). The energy of a single photon is quite large and sufficient to provide synthesis of ATP molecule or start the cascade of signalling events. Taking into account that 1) much less than 1% of photosynthetically active solar radiation is transformed to energy of photosynthetic products and 2) efficiency of photosynthesis is around 1-3% (Govindjee, 1982), plants are not ideal biochemical convertors of light. Biosphere is rather wasting solar radiation and life is not limited by energy demands, having the other intrinsic and external (e.g. water resources, carbon and nitrogen availability, temperature etc.) regulators, limitations and reasons for development.

Apart from absorbing energy of light in photosynthesis, plants also have photoreceptor systems including several phytochromes, cryptochromes and phototropins, they respond to very low intensity of different light wavelengths and have regulatory functions (Chen et al., 2004). Recently discovered plant receptor UVR8 for the range of ultraviolet-B light complements the earlier known receptors (reviewed in: Heijde, Ulm, 2012).

Light responses in photosynthesis and by photoreceptor cascades are registered by many ways and methods, from picosecond changes in spectral properties of photosynthetic pigments, conformational transitions of proteins, migration of small molecules and signalling events within and between cells, slow changes in membrane potentials of cells (Figure 1) etc. to accumulation of new pools of synthesised organic molecules-photoassimilates (Govindjee, 1982).

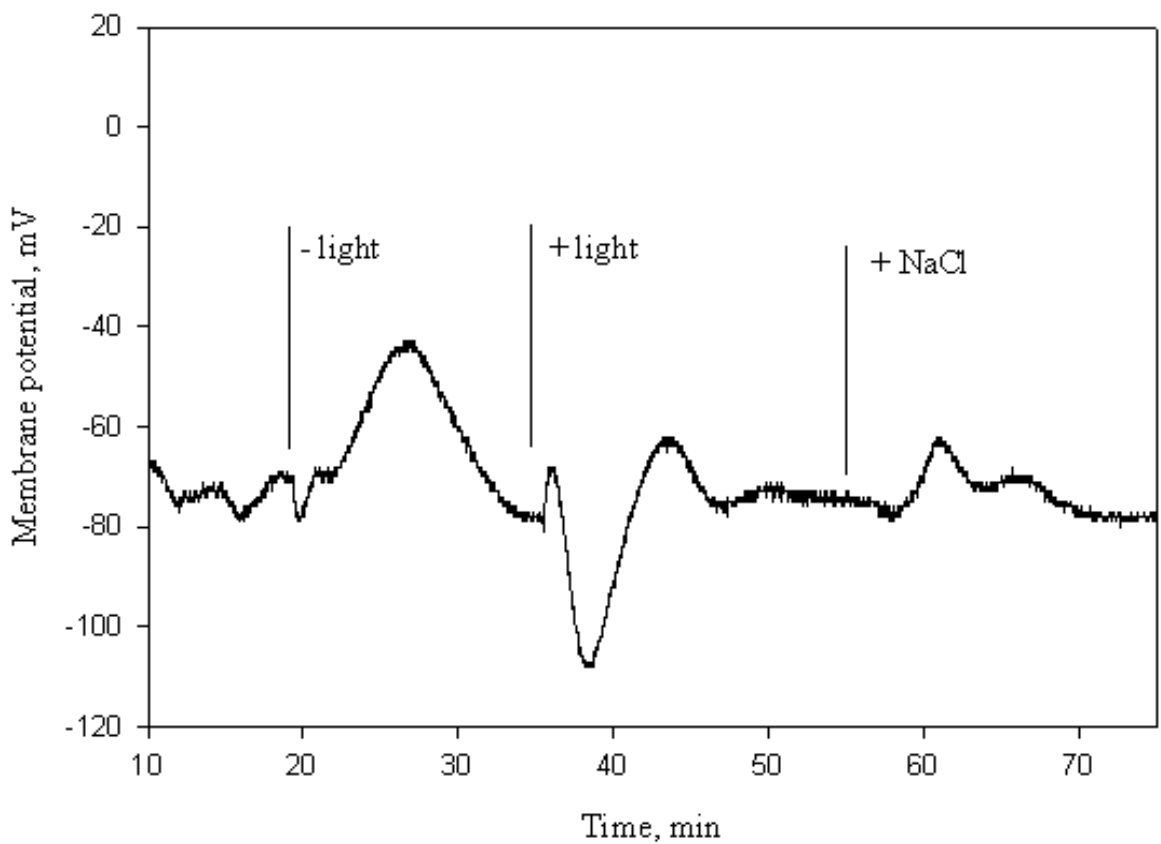

Figure 1. Typical recording of membrane potential in cells of the emerged blade of the growing leaf 3 of barley, and the response of membrane potential to changes in illumination and further addition of NaCl (100 mM) to the root medium. Several phases of responses with different kinetics were observed upon changes in illumination (18 experiments with 9 plants). Light was supplied using fibre optics from the cold light source at the background of dim illumination in the electrophysiological rig. The figure is from (Fricke et al., 2006) with permission from the Oxford University Press, extra information about the recordings is provided by V.Volkov.

Vertebrate animals, invertebrates and the other non-photosynthetic organisms get visual information using specialised organs. Eyes of vertebrates are very sensitive. The estimate for the lower limit of human eye sensitivity is about 100 photons, which will correspond to about 10-20 photons reaching photoreceptor cells due to absorption and reflection (Hecht, Shlaer, and Pirenne, 1942). Eyes contain retinal layer with photoreceptor cells, which includes 1) millions of more sensitive rods for lower illumination or night vision and may have 2) usually smaller number of less sensitive cones for colour vision (Dowling, 2012). Rods are oblong cells (Figure 2) with numerous stacked disks at the outer (distal) part; the disks are formed by membranes with photosensitive pigment rhodopsin (Dowling, 2012). The number of disks is quite large, for example about 1000 for a mouse rod; then about $10^5$ rhodopsin molecules per disk will result in nearly hundred millions of rhodopsin molecules per a mouse rod (reviewed in Palczewski, 2006).

The structure of rhodopsin is well studied: the protein part is called opsin, which is 30-50 kDa seven transmembrane G-protein coupled receptor (e.g. reviewed in: Terakita, 2005; Palczewski, 2006); the chromophore is 11-*cis*-retinal, which photoisomerises to all-*trans*-retinal after absorbing a photon (Wald, 1968). Photochemistry of rhodopsin isomerisation with picosecond rate constants and several intermediates is also well known: isomerisation has quantum yield of over 0.6 and the energy barrier is nearly 190 kJ/mole for primary reaction (energy difference is slightly over 130 kJ/mole between rhodopsin and the first photoproduct bathorhodopsin), quantum yield is about 0.1 for one of the later steps (reviewed in: Birge, 1990). Smaller values for activation energy around 100-120 kJ/mole were also reported, though there are variations between species and experimental conditions (reviewed in: Birge, Barlow, 1995).

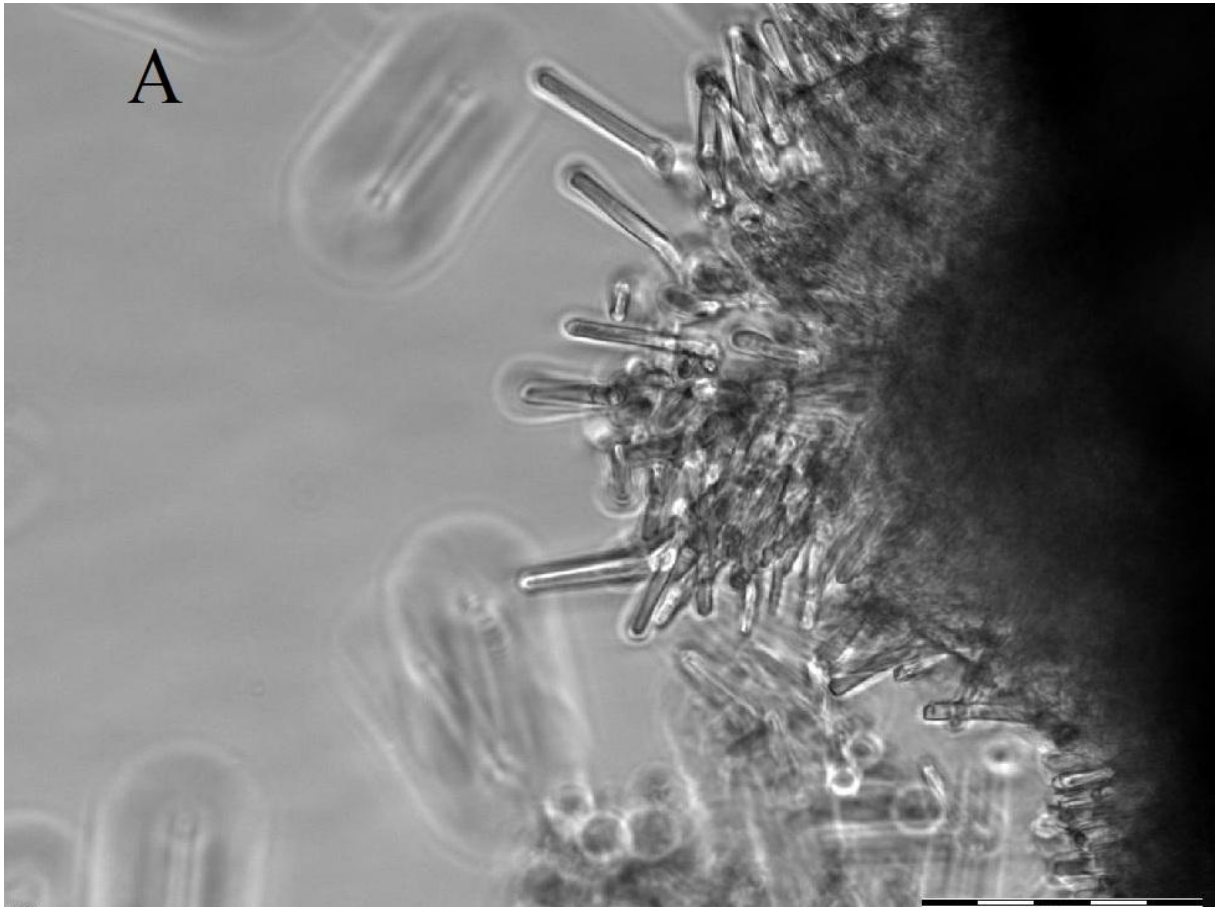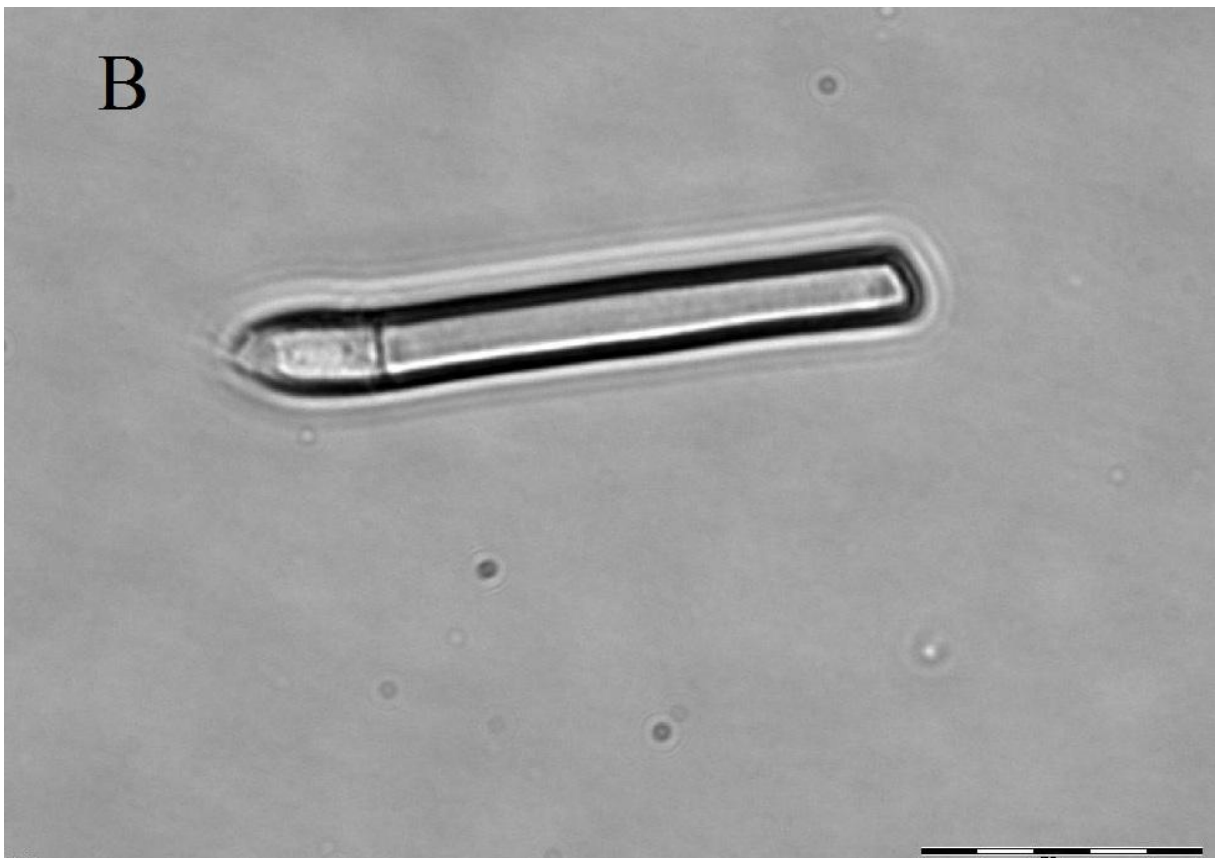

Figure 2. Microscope image of isolated *Xenopus laevis* retinal cells (A) and a rod cell from the preparation (B). Scale bar is 120 μm (A) and 30 μm (B).

Most adult insects have compound eyes, which are composed of thousands similar units called ommatidia. Each ommatidium has generally eight photoreceptor cells with rhodopsin (Land, Chittka, 2012).

For an unaided human eye the visible wavelengths are limited by 400 nm – 700 nm (Dowling, 2012), from violet to red colours. Insect eyes have slightly different wavelengths of light perception, from below 300 nm to over 700 nm (Briscoe, Chittka, 2001). The difference in UV range is explained by the secondary ultraviolet-absorbing chromophore in some insects, the sensitizing pigment can transfer the energy further to the primary photopigment (Land, Chittka, 2012).

The sequence of events from a photon hitting a rod cell to the registered photocurrent of the cell is deciphered in vertebrates (reviewed in: Stryer, 2012) (Figure 3). Rhodopsin absorbs a photon and isomerises to metarhodopsin; both proteins were crystallised and resolved structurally (Palczewski et al., 2000; Choe et al., 2011, correspondingly). Metarhodopsin has a short life half-time, so special approaches were used to crystallise this G-protein coupled receptor (Choe et al., 2011). The next step is activation of transducin. Transducin is G-protein, which is composed of α, β and γ subunits. Transducin laterally diffuses on the surface of disk membrane, interacts with activated rhodopsin (the spectroscopic intermediate metarhodopsin II) and changes bound GDP for GTP (reviewed in: Stryer, 2012; Lamb, 1996; Lamb, Pugh, 2006). Active form of transducin is α-subunit-GTP, two activated subunits bind to phosphodiestherase PDE and activate it. PDE hydrolyses cGMP to GMP and decreases concentration of the cyclic nucleotide in a cell. Sharp drop in cGMP closes cyclic nucleotide gated channels, which are regulated (gated) by bound cGMP. Under low cGMP the channels are closed, while they are in an open state in darkness under higher micromolar concentration of cGMP (Fesenko, Kolesnikov, Lyubarsky, 1985). Rod cell membrane hyperpolarises having closed cyclic nucleotide gated channels, so the initial absorption of photon expresses finally in the change of the membrane potential and corresponding ion current, which is further passed to neurons (reviewed in: Stryer, 2012; Lamb, 1996; Lamb, Pugh, 2006).

The amplification of signal is happening in the cascade: metarhodopsin R* can activate up to hundreds of G-protein transducin molecules (G) molecules (Fung, Hurley, Stryer, 1981; Stryer, 2012). Rate of activation is around 125 G* s$^{-1}$ per R* for amphibian rods at room temperature and about 3 times higher in mammalian rods at body temperature (Lamb, Pugh, 2006). Further on in darkness metarhodopsin, activated forms of tranducin and phosphodiestherase are deactivated, concentration of cGMP is increasing and membrane depolarises again; the processes add time components and kinetics to the development of the events (reviewed in: Burns, Pugh, 2010). Inactivation includes several steps, for R* in mouse rods the activity is quenched with half-time about 50-80 ms (Krispel et al., 2006; Chen et al., 2010) by successive reactions of phosphorylation by rhodopsin kinase and binding of the protein arrestin (reviewed in: Burns, Pugh, 2010).

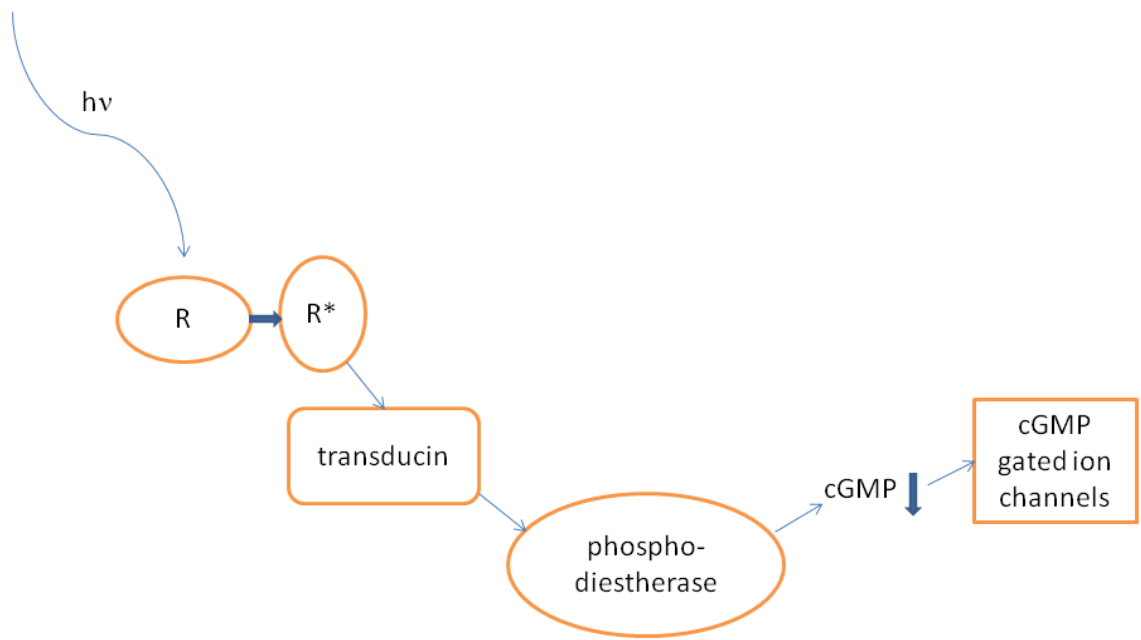

Figure 3. Signal transduction chain in a vertebrate retinal rod starting from a photon hν and leading to cyclic nucleotide gated channels. Closure of the ion channels after drop in cGMP results in membrane hyperpolarisation and stops inward ion current of sodium and calcium.

The kinetics of induced photocurrent in rods is reasonably modelled by several equations, which take into account kinetics of the reactions (rhodopsin to metarhodopsin, activation of transducin, phosphodiestherase, drop in cGMP and closure of ion channels), include diffusion, rod morphology and inactivation kinetics (eg: Cobbs, Pugh, 1987; Burns, Pugh, 2010). Higher number of photons increases the photocurrent activating more rhodopsin molecules; the response is nonlinear and reaches saturation at around eg 20,000 photons per light pulse for *Xenopus* rod (Sim et al, 2012). Numerous mutations affecting components of the signal transduction chain for photon perception are known, some of them have an effect on photocurrent and its' kinetics (reviewed in: Burns, Pugh, 2010). Surprisingly stable and reproducible kinetics of photocurrent upon a given number of photons provides robust and reliable information about the light source. The rod photoreceptor could be considered like a natural example of engineering with numerous feedbacks; inactivation components are especially important for ensuring reproducible responses (Rieke, Baylor, 1998). For example, C-terminus of rhodopsin has 6 phosphorylation sites, which are important for inactivation of activated R* and reproducible kinetics of photocurrent; decreasing the number of the sites in mouse mutants increased the duration of photocurrent and changed its' shape (Doan et al., 2006).

Invertebrates may have slightly different sequence of events during phototransduction. Fruitfly *Drosophila* is a well-known model in genetics; numerous mutants are useful for deciphering the light perception in the ommatidia of the insect. The difference in phototransduction from vertebrates is that in *Drosophila* 1) phospholipase C is present instead of phosphodiestherase, 2) signalling via inositol trisphosphate and diacylglycerol and probably polyunsaturated fatty acids without cGMP 3) results in opening of closed under darkness 4) transient receptor potential ion channels (reviewed in: Montell, 1999, 2012). More differences could be revealed among species of numerous and strikingly unusual biological organisms.

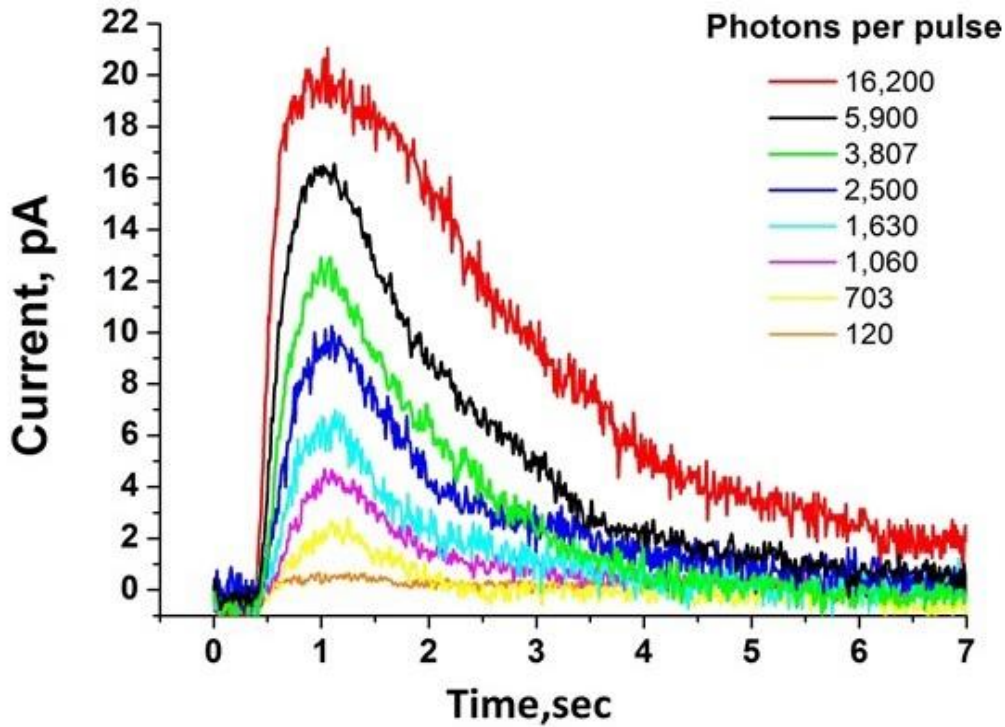

Figure 4. Kinetics of retinal rod responses of *Xenopus laevis* to pulses of illumination. The colour legend indicates the corresponding number of impinging photons. Reproduced with the permission of The Optical Society of America from (Sim et al., 2011).

Obviously, the ion current is the only converted measure of light intensity and it is further transferred to neurons of the neural system. Artificial electronic light receptors were proposed instead of absent or damaged retinal cells under ophthalmological diseases; the electric response of the retinal prosthesis is passed to neurons restoring (at least partially) light sensing (reviewed in: Chader, Weiland, Humayun, 2009; Nirenberg, Pandarinath, 2012). Compared to retinal photoreceptor cells, which are highly specialised and finely tuned systems, retinal prostheses are simpler. Another approach is to express light-sensitive proteins in neurons. Several genes from *Drosophila* including rhodopsin rendered light sensitivity to hippocampal neurons in primary culture (Zemelman et al., 2002). Directly activated by light ion channel channelrhodopsin is also proposed for use in retinal prosthesis

(Nirenberg, Pandarinath, 2012). Model experiments using mouse model with degenerated retinas demonstrated that being genetically targeted to inner retinal neurons channelrhodopsin-2 can restore basic visual function with, however, 6-9 orders of magnitude lower sensitivity (Bi et al., 2006; Lagali et al., 2008) (Figure 5).

Here emerges an interesting area of discrimination between simple photic responses (difference light-dark and corresponding physiological reactions) and recognition of images derived from visual perception by specialised photoreceptor cells. The direction leads to neuroscience with different specialised types of neurons and is outside the scope of the present review. For example, photosensitive retinal ganglion cells were found in mice, the cells express melanopsin and confer simple photic responses in blind mice without photoreceptor rods and cones (Panda et al., 2003). Human melanopsin from intrinsically photosensitive retinal ganglion cells was heterologously expressed in mouse paraneuronal cell line Neuro-2a and rendered light-induced ion currents (Melyan et al., 2005). Mouse melanopsin was heterologously expressed in *Xenopus* oocytes (Panda et al., 2005) and in human embryonic kidney HEK293 cells (Qiu et al., 2005) making them light-sensitive. Visual perception is much more complicated.

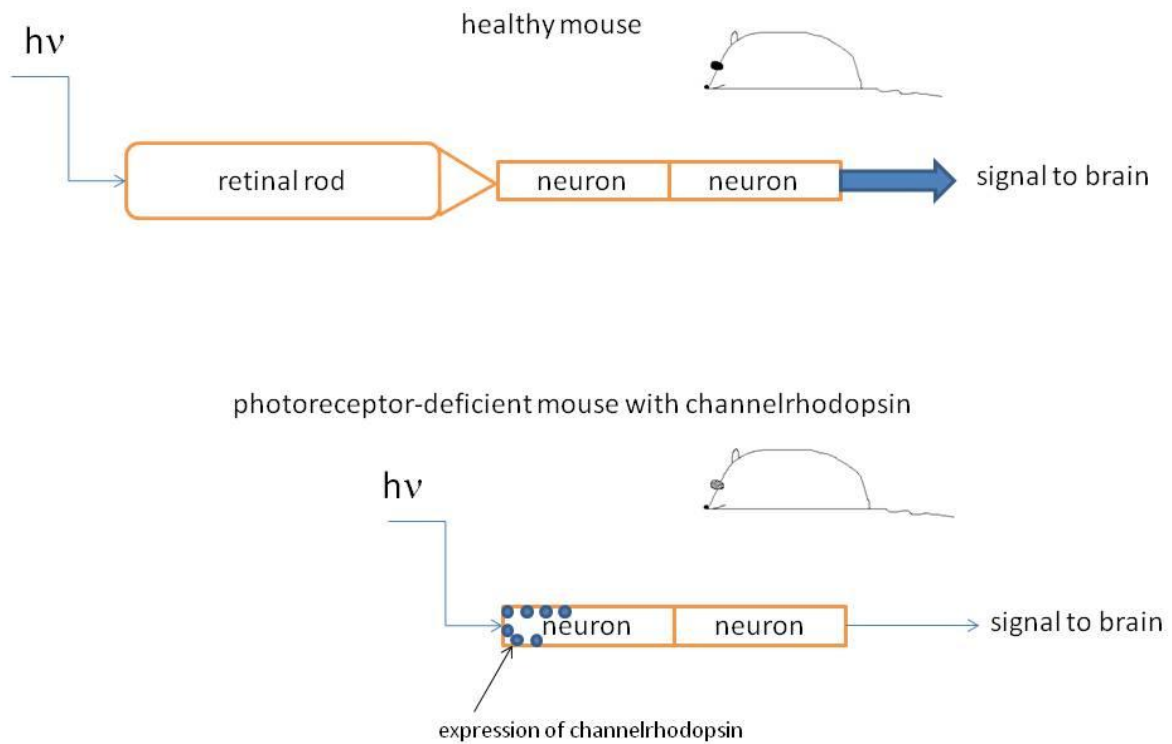

Figure 5. Light-activated ion channel channelrhodopsin-2 can restore basic visual function in photoreceptor-deficient mice with 6-9 orders of magnitude lower sensitivity (according to (Bi et al., 2006; Lagali et al., 2008)).

Since each biological system is a physico-chemical system, it can be described in terms of energies, activation energies, kinetics of reactions and the other parameters. Numerous restrictions are implied by biological components (nature of interacting molecules; morphology of cells and tissues; pH, redox potential and chemical composition of cell medium). On the opposite, obvious predictions about behaviour of biological systems could be done from the known parameters. The simplest prediction is about rhodopsin isomerisation. High activation energy of photoisomerisation for rhodopsin can not prevent it from spontaneous isomerisation without any photons, but due to temperature (around 300 K) and uneven distribution of energies of interacting molecules. One reason of so called dark noise of photoreceptor appears, it will depend (for the reason) on temperature and number of rhodopsin molecules in a rod cell. Dark noise of photoreceptor is expressed in spontaneous electric signals (voltage spikes or recorded ion currents), which are not distinguishable from the electric signals induced by photons. Indeed, dark noise of photoreceptor measured in events/(cell*s) (varied from $10^{-3}$ to 1) was linearly proportional to the number of rhodopsin molecules within the range of $10^8$ - $10^{10}$ rhodopsin molecules/cell (locust-human-toad-*Limulus*) (Birge, Barlow, 1995), means around $10^{-10}$ events/second per a rhodopsin molecule. However, the situation looks much more complicated taking into account the sequence of events from activated rhodopsin to closure of ion channels. Each element (spontaneous activation of transducin or phosphodiestherase, drop in cGMP or stochastic closure of ion channels) has to be assessed using activation energies and stoichiometry of reactions.

An interesting observation was done for photoreceptor cells of horseshoe crab *Limulus*: dark noise was about one bump of voltage per 5 minutes during evening and 25 times higher during midday (Barlow et al., 1997). ). On the opposite, the voltage response to illumination (gain) was much higher during night time. The model of two-step process for spontaneous temperature-dependent isomerisation of rhodopsin was proposed (Barlow et al., 1993; Birge, Barlow, 1995). The first step is deprotonation of protein-bound chromophore (form of retinal) with activation energy around 100 kJ/mole. The second step is isomerisation of the chromophore with activation energy around the remaining 100 kJ/mole. The first step depends on pH and geometry of protein-chromophore complex, so changes in pH and protein microenvironment can strongly influence the noise. Temperature-dependent isomerisation of chromophore at the second step can account for the observed temperature dependence of noise, which was the same during daytime and during night. Indeed, for photoreceptor cells of horseshoe crab the noise increased about twice with temperature increase by $6^0$C (from 12 events/s at $14^0$C to 28 events at $20^0$C during daytime and from 3 events/s at $20^0$C to 6.5 events/s at $26^0$C during night) giving the estimated energy activation of 90-100 kJ/mole (Barlow et al., 1993).

However, the origin of photoreceptor noise and temperature dependence of photoreception could be much more complex. Dependence of thermal noise on pH of external medium was found for photoreceptor cells of horseshoe crab (Barlow et al., 1993), but not for cones of salamander (Sampath, Baylor, 2002) or toad rods (Firsov, Donner, Govardovskii, 2002). Noise in salamander cones (Rieke, Baylor, 2000) had different origin (due to pigment or due to transduction species) for different types of cones.

Finally, physiological mechanisms and biological peculiarities are important for temperature-dependence of visual perception. For example, in swordfish (*Xiphias gladius*) the temporal resolution of retinograms (electrical activity recorded from isolated retinal preparations) was measured by the flicker fusion frequency. The parameter determines the frequency of pulses, which are distinct in retinograms. The flicker fusion frequency has very high temperature coefficient $Q_{10}$ (change of measured parameter with changing temperature by $10^0$C) in swordfish: about 5; resolved frequency is around 30 Hz at $21^0$C and just 6 Hz at $6^0$C (Fritsches, Brill, Warrant, 2005). Swordfish has twice higher $Q_{10}$ of flicker fusion frequency than bigeye and yellowfin tunas, but possesses special system for heating brain and eyes and, therefore, keeps good visual perception at depths of several hundred meters (Fritsches, Brill, Warrant, 2005).

At the molecular level it's reasonable to start from the point that a clear model for thermal isomerisation of rhodopsin still has to be proposed, especially when rhodopsin is located in membranes of rod outer segments and interacts with lipids, water molecules and proteins. For comparison, energy of gas molecules is described by Maxwell-Boltzmann distribution and statistics and in a simplified form can be expressed by:

$$N_i/N = e^{-\Delta E/kT},$$

where N is the total number of particles; $N_i$ is the number of particles with energy by $\Delta E$ higher than a certain energy (usually probabilities $N_i/N$ are considered); Bolzmann constant k = $1.38*10^{-23}$ J/K, T is temperature (e.g. Chang, 1977, p.203).

A simple estimate is useful to make an analogy to gas systems with the corresponding parameters: 1) the temperature is 300 K ($27^0$ C), 2) activation energy $E_a$ equals to 100 kJ/mole and for the known Avogardo number Na = $6.02*10^{23}$ molecules/mole gives $E_a/N_a$ = 100 000 Joles /$6.02*10^{23}$ = $1.6*10^{-19}$ Joles/molecule, then part of gas molecules with energy higher the energy barrier of $1.6*10^{-19}$ Joles/molecule will be: exp(-$1.6*10^{-19}$/($1.38*10^{-23}$*300)) ≈ exp(- 39) ≈ $3*10^{-17}$. Obviously, the number is far below the reasonable values of the observed dark noise ($10^{-10}$ events/second per a rhodopsin molecule for $10^{10}$ molecules per a photoreceptor rod while the activated rhodopsin molecule has a life half-time of about 50 ms); it proves that quantum effects of photochemical processes and interactions with surrounding molecules are involved to change the distribution.

Pondering the smart and ideal biological design, it's reasonable to think about photoreceptor without any dark noise. However, the biological systems are too noisy and have relatively high level of errors compared to technical or digital computer systems. Potentially the noise could be beneficial for the visual perception keeping a certain background activity of neurons. More general considerations propose limits for noise suppression in biological systems, otherwise negative feedbacks and the whole functioning of system are getting extremely expensive: the minimum standard deviation decreases with the quartic root of the number of events for Poisson communication channels (Lestas, Vinnicombe, Paulsson, 2010).

Polarised light, occurrence in nature and mechanisms of perception

Apart from wavelength and intensity electromagnetic waves have polarisation. Electromagnetic wave consists of fluctuating electric and magnetic fields, which have perpendicular orientation to each other. Considering a single photon like a wave packet and a pulse of light with thousands and millions of photons, which are not ordered and have random directions of individual vectors of electric field, it's simple to understand unpolarised light. However, depending on the light source and passed medium the orientation of electric vector in the wave may become ordered, the electromagnetic wave is becoming polarised. If the orientation of electric vector fluctuates in one plane, then the wave is linearly polarised; alternatively the vector can rotate and the wave is circularly polarised then. The degree of polarisation characterises the extent of ordered orientation of the electric vector. Polarisation of light occurs naturally in atmosphere during sunrise or sunset, after reflection of light from water surface, from Moon and also underwater (Können, 1985; Pye, 2010; Cronin, Marshall, 2011). So, the polarised light has essential information for biological organisms about time and surrounding environment. Unfortunately, it can not be directly digitised and translated to a fixed digital image.

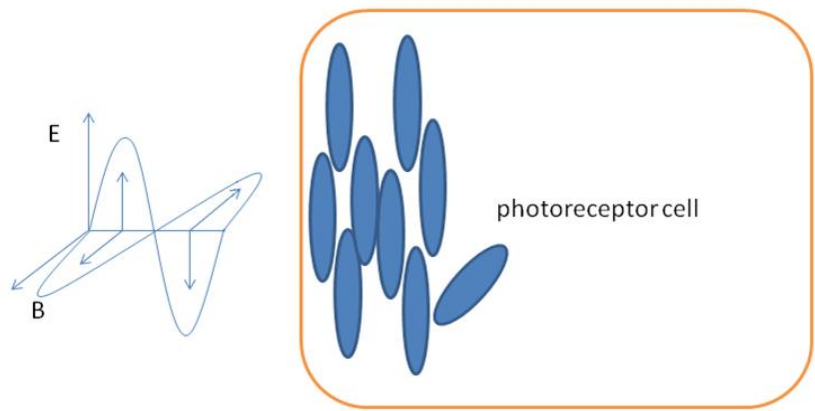

Figure 6. Basic principle for perception of polarised light in biological organisms. Ordered orientation of pigment molecules in photoreceptor cell allows detection of electric vector of light

Human beings have low sensitivity for polarised light and able to detect just very high polarisation degrees, over 60%, while some insects and fishes can see polarisation of 10% even (Können, 1985). Polarisation of light is more important for invertebrates and fishes due to their habitat and size. Some insects are able to use polarised light for behavioural signalling. The metallic green elytra (forewings) of Japanese jewel beetle *Chrysochroa fulgidissima* reflects light, which is becoming highly polarised because of complex multi-layered surface of the elytra (Stavenga et al., 2011). South American beetle *Plusiotis boucardi* with metallic-looking forewings has dual-pitched helicoidal layer with bowl-shaped recesses, the forewings are special in reflecting circularly polarised light without changing its handedness (Jewell, Vukusic, Roberts, 2007). The degree of polarisation may depend on the angle of reflection and the polarised light reflected by the beetles may have function in recognition of the species. In butterfly *Heliconius* the reflected polarised light from the wings can be a preferential signal for mate recognition (Sweeney, Jiggins, Johnsen, 2003).

On the opposite, in fish and underwater species polarisation pattern of light reflected from the skin or epidermal tissues can be more important for being less visible by predators. Squid has iridescent skin with multilayer reflector cells; the cells contain plates of protein interspersed by cytoplasm and allow the animal to change colour and polarisation of reflected light and to send potential polarisation signals to the other cephalopods with polarisation-sensitive vision (Moody, Parriss, 1960; Mäthger, Hanlon, 2006; Mäthger, Shashar, Hanlon, 2009). Several species of fish (sprat *Sprattus sprattus,* sardine *Sardina pilchardus*, and herring *Clupea harengus*) have similar cell structures: broadband guanine-cytoplasm 'silver' multilayer reflectors, which make, however, the reflected light non-polarised and provide a sort of cryptic camouflage for the situation (Jordan, Partridge, Roberts, 2012). The study on the structure of insect and animal polarisation surfaces (wings, forewings, skin etc.) may have important implications for material science.

The perception of polarised light, obviously, requires ordered spatial positioning of rhodopsin/visual pigments and specialised cells for perceiving the direction and degree of polarisation. Much more is known about behavioural reactions of different species to polarised light than about mechanisms of perception of the light, though a few detailed studies on the perception exist and multiply. In continuation of the earlier experiments describing polarisation-dependent dances of bees (Frisch, 1949), it was shown that bees possess receptors of polarised ultraviolet light, but not of the other wavelengths (Helversen, Edrich, 1974; Sakura, Okada, Aonuma, 2012). The specialised receptors at the dorsal rim of bee's compound eyes were examined electrophysiologically (Labhart, 1980). Very high polarisation sensitivity up to 18 (ratio of maximal to minimal voltage responses to several directions of electric vector of light) was found in UV-responsive cells lacking sensitivity to green light (Labhart, 1980). In fish the perception of ultraviolet polarised light is associated with cones (Hawryshyn, 2000), while the polarised light of another parts of spectrum is also sensed according to electrical recordings from brain tectum (Waterman, Aoki, 1974: Waterman, Hashimoto, 1974).

The ultrastructure of photoreceptors for polarised light in arthropods is studied for many species (eg. and reviews: Nilsson, Labhart, Meyer, 1987; Labhart, Meyer, 2002; Mueller, Labhart, 2010; Roberts, Porter, Cronin, 2011). Usually the pigments are packed orderly in microvilli within a photoreceptor cell; in each ommatidium the direction of microvilli is orthogonal (at $90^0$) in the photoreceptor cells. So, the direction of electric vector of light can be detected by different cells, the signal is periodically sent to neurons, which process the information about polarisation. Periodical signal from different cells helps to exclude effect of light intensity, while the initial polarisation of light is often amplified within an ommatidium or an arthropod eye by reflecting surfaces (reviewed in: Labhart, Meyer, 2002; Mueller, Labhart, 2010; Roberts, Porter, Cronin, 2011). New approaches including methods of molecular biology and genetics help in dissecting and understanding the components of the polarisation vision (Wernet et al., 2012).

It looks challenging to mix 1) perception of electric vector of polarised light with the electric sense of several sea inhabitants including sharks and rays (Kalmijn, 1971; Antipin, Krylov, Cherepnov, 1984; Gonzalez, 2008) and 2) magnetic vector of polarised light with magnetoreception of many living organisms (eg reviewed in: Johnsen, Lohmann, 2005). The electric sense of sharks and rays has resolution below 5 nV/cm (Kalmijn, 1971; Antipin, Krylov, Cherepnov, 1984; Gonzalez, 2008) and seems comparable or even much lower than the electric field in light beams: eg. sunlight has intensity of electric vector around 10 V/cm (Singh, 2008, p.10). However, the distinct sensual informational channels are determined by:
1) range of measured values and frequencies of electromagnetic fields (hundreds of Hz for electric sense and around $10^{14}$ - $10^{15}$ Hz for visible light);
2) detailed possible mechanisms (often not studied yet) of perception;
2) morphological structures for specific sort of perception that are based on the possible mechanisms created and selected by nature (specialised photoreceptor cell types in eyes, ampullae of Lorenzini for electroreception in sharks and rays, magnetite-containing cells for magnetoperception etc.);
4) further numerous peculiarities.

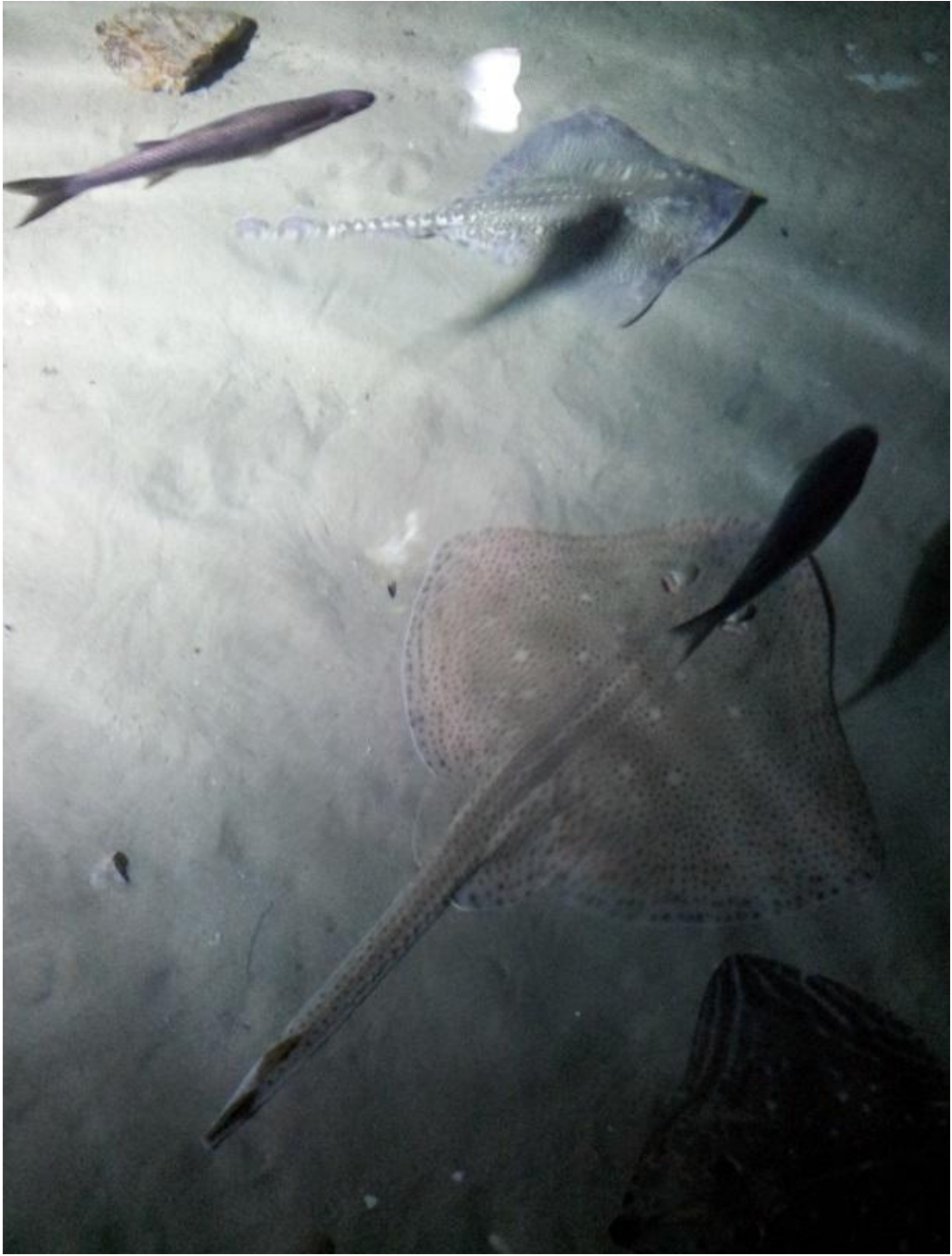

Figure 7. Underwater world with several photobiological phenomena and examples of perception. Light is reflected from the water surface and becoming partially polarised; light is also passing to water and getting partially linearly polarised; polarised light in water is sensed by fishes; fishes have multilayer reflecting skin with unusual properties; blonde, thorntail and undulate rays (*Raja brachyura*, *Raja clavata*, *Raja undulate*; size of specimens is around 1 m) possess electric sense.
The picture is taken with the permission of staff of SEA LIFE London Aquarium.

Polarised light and direction of polarisation may have an effect on development of biological organisms. For example, protonemata (primary germinating structures in development from spores) of ferns and mosses have preferential orientation according to electric vector of illumination (Bünning, Etzold, 1958; Nebel, 1969; Jenkins, Cove, 1983a, b; Kadota et al., 1984; Jaedicke, 2012). Interesting experiments were carried out with isolated and immobilised in agar protoplasts (single cells obtained from plants using cell wall degrading enzymes) of moss *Physcomitrella patens* (Skripnikov, Shefer and Zrid, 2009). The phragmoplast (scaffold for future cell wall between the new forming cells) of the first cell division was perpendicular to the electric vector of polarised light: nearly 50% of protoplasts had phragmoplasts at angles $80^0$-$110^0$ to electric vector compared to around 5% at e.g. angles $140^0$-$170^0$. The polarisation was induced by covering protoplasts in Petri plates by plastic polarisation filters, no ordered orientation of phragmoplasts was found at non polarised light (Skripnikov, Shefer and Zrid, 2009). Phytochorome signalling is involved in the responses since effects of red light were reverted by far red light or were synergetic (Nebel, 1969; Kadota et al., 1984). No effects of polarised ultraviolet or blue light were found for apical growth of fern *Adiantum capillus-veneris* (Kadota et al., 1984), but the effects of polarised blue and green light were present for growth of moss *Physcomitrella patens* (Jenkins, Cove, 1983a). Obviously, ordered localisation of phytochorome (and probably of the other photoreceptors) is required for the reaction, which may also imply cytoskeleton reorganisation and calcium waves/signalling.

Effects of different statistics of photons on light perception

The main source of light existing in nature is solar radiation with properties of thermal light. Photons are emitted from numerous energy levels and the distribution of their number in a first approximation obeys Bose-Einstein statistics. The other well-known sources of light with different parameters are lasers. Distribution of photons emitted by lasers obeys Poisson statistics. Lasers provide a valuable tool for biological research and are widely used in technique due to their unique features (high possible power per a pulse, monochromatic light, high degree of polarisation, low noise etc.). One of simple differences between the two statistics is that number of photons per unit of time is more uniform for lasers; functions for probabilities of distribution of photon number are different for thermal light and for lasers (Mandel, Wolf, 1995).

Randomly occurring fluorescence, light from fires etc. usually are not discussed since having no known information/predictable effects for biological objects. Bioluminescence is widely spread (e.g. reviewed in: Haddock, Moline, Case, 2010) though doesn't look different from the point of photon statistics and polarisation. There are reports about super-Poisson statistics (typical for thermal light) of bacterial (*Photobacterium phosphoreum*) bioluminescence (Kobayashi, Devaraj and Inaba, 1998) and photon emission from cellular slime mold (*Dictyostelium discoideum*) during developmental processes (Kobayashi and Inaba, 2000), which, however, need further investigation and confirmation.

An interesting question appears whether different sources of light may have different effects on biological systems due to statistics of number of photons. Potentially the same number of photons with the same energy, but distributed differently within the same time of illumination pulse may result in different effects. For example, conformations of proteins are subject to fast reversible and irreversible fluctuations due to thermal noise, interactions, changing microenvironment (at the scale of a few nanometers). The estimated upper limit for frequency of protein conformational changes is around $10^6$ Hz (Chakrapani, Auerbach, 2005). Assuming just interaction of photons with one protein, it's conceivable to imagine several conformational levels for a protein and different effects after a multiphoton pulse of thermal or laser illumination.

A few experimental papers describe comparison of different sorts of illumination on visual perception paying attention to the statistics of photons (eg: Teich et al., 1982a, 1982b). Interesting results were obtained with retinal rods of *Xenopus*: the slope of response to Nd:YAG laser at 532 nm was steeper than for pseudothermal (the same statistics like thermal) light (Sim et al., 2012). Photocurrent response of *Xenopus* rod cells was saturated by about 25,000 photons per a 30 ms pulse in the experiments; the difference between laser and pseudothermal light appeared after half-saturating amplitude of photocurrent. Relative photocurrent (normalised to saturating) was about 30-40% higher for laser source of photons in the range of illumination (Sim et al., 2012). The observation is most likely connected to transduction chain species (the life of activated rhodopsin is about 50-80 ms) and raises questions for most experiments with laser light. It might be possible that results with lasers and Poisson statistics require corrections when approximating for thermal sources of light.

The future directions in photobiology are bright and spread far outside the scope of the review. Evident progress of optogenetics is expressed nowadays in potential medical applications. Further and deeper understanding of photobiological processes including leap to spatial nanoscale and temporal femtoscale in combination with new approaches of molecular biology and genetics needs also integrative and synthetic way of seeing. The new and more detailed picture with higher resolution will rise. More knowledge is gained from different species, so details of phototransduction may vary and leave plenty of space for future research.

Acknowledgement. Unfortunately, I can not number and thank all colleagues who participated in numerous discussions within over 20 years of my fluctuating interest in the area, who inspired some ideas and thoughts, so I can just apologise for not citing all the relevant literature sources.